\documentclass[conference, a4paper]{IEEEtran}
\IEEEoverridecommandlockouts
\usepackage{cite}
\usepackage{amsmath,amssymb,amsfonts}
\usepackage{amsthm}
\usepackage{bm}
\usepackage{algorithmic}
\usepackage{graphicx}
\usepackage{textcomp}
\usepackage{xcolor}
\usepackage{mathtools}
\usepackage{threeparttable}
\usepackage{subfig}
\usepackage{float}
\usepackage{overpic}
\usepackage{pifont}
\usepackage{color}
\usepackage{algorithm}
\usepackage{hyperref}
\usepackage{cleveref}
\usepackage{booktabs}
\usepackage{multirow}
\usepackage{makecell}

\usepackage[
top    = 0.8in,
bottom = 0.95in,
left   = 0.52in,
right  = 0.52in]{geometry}

\def\BibTeX{{\rm B\kern-.05em{\sc i\kern-.025em b}\kern-.08em
    T\kern-.1667em\lower.7ex\hbox{E}\kern-.125emX}}
\begin{document}

\title{AEPHORA: AI/ML-Based Energy-Efficient Proactive Handover and Resource Allocation}

\author{\IEEEauthorblockN{
Bowen Xie\IEEEauthorrefmark{1}, 
Sheng Zhou\IEEEauthorrefmark{1},
Zhisheng Niu\IEEEauthorrefmark{1},
Hao Wu \IEEEauthorrefmark{2},
Cong Shi \IEEEauthorrefmark{2}}\\
 \IEEEauthorblockA{\IEEEauthorrefmark{1}Beijing National Research Center for Information Science and Technology\\
  Department of Electronic Engineering, Tsinghua University, Beijing 100084, China\\
  \IEEEauthorrefmark{2}Standard Research Department, OPPO, Beijing, China\\
  Email: xbw22@mails.tsinghua.edu.cn, 
   \{sheng.zhou, niuzhs\}@tsinghua.edu.cn, 
   \{wuhao8, shicong\}@oppo.com}\\
 \thanks{The work of B. Xie, S. Zhou and Z. Niu are sponsored in part by the National Key R\&D Program of China No. 2020YFB1806605, by the National Natural Science Foundation of China under Grants 62022049, 62111530197, and by Beijing OPPO Telecommunications Corp., Ltd.}
}

\maketitle

\begin{abstract}

Future Vehicle-to-Everything (V2X) scenarios require high-speed, low-latency, and ultra-reliable communication services, particularly for applications such as autonomous driving and in-vehicle infotainment. Dense heterogeneous cellular networks, which incorporate both macro and micro base stations, can effectively address these demands. However, they introduce more frequent handovers and higher energy consumption. Proactive handover (PHO) mechanisms can significantly reduce handover delays and failure rates caused by frequent handovers, especially with the mobility prediction capabilities enhanced by artificial intelligence and machine learning (AI/ML) technologies. Nonetheless, the energy-efficient joint optimization of PHO and resource allocation (RA) remains underexplored. In this paper, we propose the AEPHORA framework, which leverages AI/ML-based predictions of vehicular mobility to jointly optimize PHO and RA decisions. This framework aims to minimize the average system transmission power while satisfying quality of service (QoS) constraints on communication delay and reliability. Simulation results demonstrate the effectiveness of the AEPHORA framework in balancing energy efficiency with QoS requirements in high-demand V2X environments.
  
\end{abstract}

\section{Introduction}

Modern vehicles are transforming into intelligent mobile terminals, necessitating real-time and efficient communication with their environments, other vehicles, and pedestrians. This evolution has led to the emergence of intelligent connected vehicles (ICVs) and Vehicle-to-Everything (V2X) communication\cite{c-v2x}. ICVs leverage V2X technology to connect with infrastructure, pedestrians, and the Internet, supporting applications such as autonomous driving, intelligent traffic management, and in-vehicle entertainment, all of which demand differentiated high quality of service (QoS). To meet the diverse communication QoS requirements in high-mobility scenarios, the heterogeneous networks (HetNets) with macro and micro base stations (BSs) is crucial for enhancing vehicular network performance\cite{andreev2019future}. This coordinated deployment strategy ensures large signal coverage in V2X scenarios while significantly lowering overall system energy consumption and data service latency\cite{xu2021survey}.

However, V2X communication still faces several challenges. The high-speed movement of vehicles, coupled  with the dense deployment of BSs, leads to frequent handovers (HOs) and associated ping-pong effects, resulting in increased signaling overhead and degraded communication quality \cite{hasan2018frequent}. Traditional LTE HO schemes experience higher HO failure (HOF) rates and longer HO latency in dynamic environments, particularly in the millimeter-wave (mmWave) band \cite{lee2020prediction}. Moreover, energy-efficient HO and resource allocation (RA) schemes are in urgent need\cite{manalastas2020go}.

Existing studies propose the proactive handover (PHO) strategy to address these issues \cite{lee2020prediction,sun2018energy,park2024mobility}, which involves forecasting traffic flow, vehicle trajectories, and wireless channel states to prepare HO in advance. By leveraging artificial intelligence and machine learning (AI/ML) technologies, it becomes feasible to estimate future vehicle positions or identify the optimal target BS based on historical channel state information\cite{park2024mobility,10155554}. This predictive capability enhances PHO by reducing HO latency, HOF rates, and ping-pong effects. However, these studies do not consider the coordinated optimization of the PHO and RA strategies to further minimize system power while meeting the communication delay and reliability requirements.

This paper investigates the PHO and RA problems in V2X networks covered by macro and micro BSs. In this context, the PHO mechanism is employed with an AI/ML-based mobility prediction method to forecast the future positions and channel states for vehicles. We propose an energy-efficient PHO algorithm and a heuristic RA algorithm that jointly optimize PHO and RA decisions to minimize the average system power while satisfying the ultra-reliable and low-latency communication (uRLLC) QoS constraints. These proposed algorithms form the AEPHORA framework. It can adaptively offload vehicles from overloaded BSs to those with sufficient spectrum resources when the traffic loads are heavy. Simulation results demonstrate the effectiveness of the proposed AEPHORA framework in achieving a good balance between system power and reliability.

\section{System Model}{\label{sec:System Model}}

\begin{figure}
\centerline{\includegraphics[width=0.45\textwidth, trim= 60 200 300 50,clip]{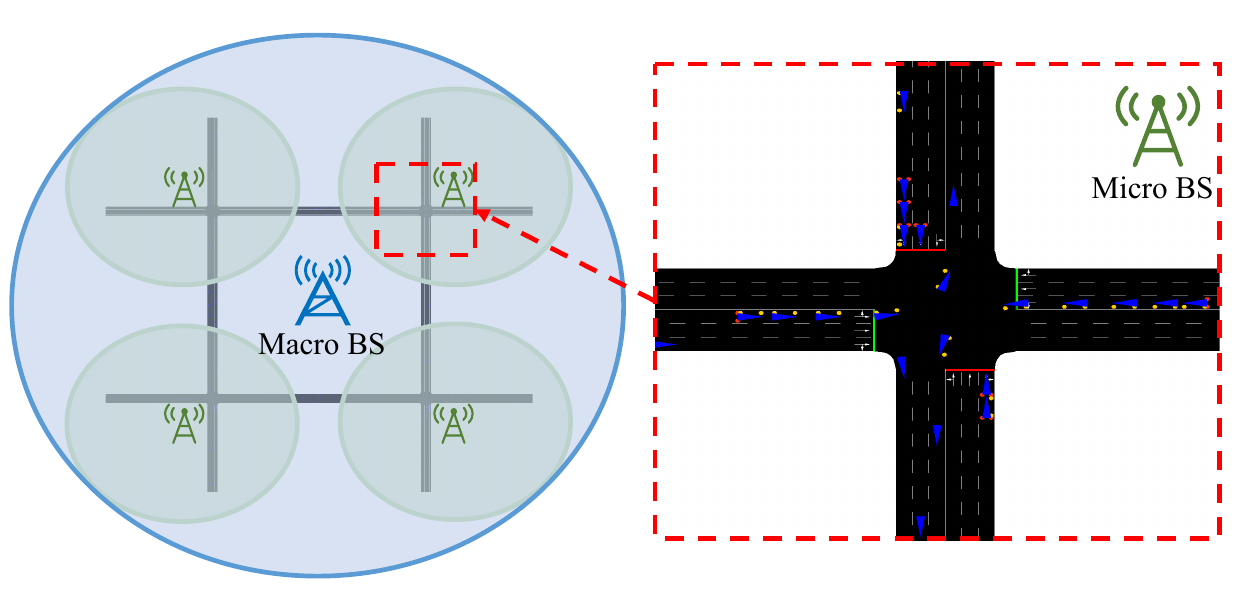}}
\caption{The system model.}
\label{fig:system model}
\end{figure}

The paper considers a vehicular network downlink communication scenario with coordinated coverage by macro and micro BSs, as illustrated in Figure \ref{fig:system model}. In this scenario, there is one macro BS, $M$ micro BSs (in Figure \ref{fig:system model}, $M=4$) and a varying number of vehicles in the considered region $\mathcal{R}$. 
Each BS is equipped with $N_{{\rm TX}}$ transmit antennas, whereas each vehicle is equipped with a single receive antenna, thereby forming a multiple-input single-output (MISO) system. 
The BS set is denoted as $\mathcal{M}=\{ 0,1,\dots,M \}$, where BS-$m$ represents the macro BS for $m=0$ and a micro BS for $m > 0$.
BS-$m$ operates in the frequency band $f_m$ and is allocated $K_m$ resource blocks (RBs), with each RB utilizing a bandwidth of $\Delta f_m$ and a transmission power of $p_m$. It is assumed that all micro BSs share the same parameters, i.e., $f_m = f_1$, $K_m = K_1$, $\Delta f_m = \Delta f_1$, and $p_m = p_1$ for all $m > 0$.

The system operates in discrete time, with the time slot length denoted by $\Delta t$. In the $n$-th time slot, the set of vehicles in the network is denoted by $\mathcal{V}(n)$, and the number of vehicles is $V(n)$. The binary variable $u_{m,v}(n) \in \{0,1\}$ represents the connection state between BS-$m$ and vehicle-$v$ in time slot-$n$, where $u_{m,v}(n) = 1$ denotes an active connection and $u_{m,v} = 0$ indicates no connection. Assume that each vehicle can only maintain a connection with one BS at a time, hence we have
\begin{equation}
    \sum_{m=0}^{M} u_{m,v}(n) \leq 1.
    \label{eq:connection cons}
\end{equation}
The connection variable $u_{m,v}(n)$ is periodically updated by the Proactive Handover (PHO) process, with an update cycle of $N_{\rm HO}$ time slots, referred to as a PHO frame. A PHO decision made in the $(x-1)$-th PHO frame takes effect at the start of the $x$-th PHO frame, and the corresponding $u_{m,v}(n)$ remains unchanged throughout the $x$-th PHO frame. 
For the sake of notational simplicity, we denote $\{\cdot\}^{[x]} = \{\cdot\}(xN_{{\rm HO}})$, $\{\cdot\}^{[x]}(i) = \{\cdot\}(xN_{{\rm HO}} + i)$, $\mathcal{N}_{{\rm HO}} = \{0,1,\dots,N_{{\rm HO}}-1\}$.
Thus, we have
\begin{equation}
    u_{m,v}^{[x]}(i)=u_{m,v}^{[x]}, \forall i \in \mathcal{N}_{{\rm HO}}.
    \label{eq:PHO cons}
\end{equation}

\textcolor{black}{The macro BS is deployed at the center of the vehicular network area to ensure comprehensive signal coverage across the entire region. The micro BSs are deployed at high-traffic intersections or road segments to enhance network capacity and communication efficiency. Additionally, the macro BS operates in the sub-6 GHz band, where each RB has smaller bandwidth but higher transmission power. In contrast, the micro BSs operate in the mmWave band, offering larger bandwidth per RB with lower transmission power. Therefore, we have $f_0 < f_1$, $\Delta f_0 < \Delta f_1$ and $p_0 > p_1$. 
Given the wired connections between the macro and micro BSs, which facilitate efficient data exchange and signaling, it is assumed that all BSs share channel and context information, enabling centralized PHO and RA decisions.}

\subsection{Communication Model}

Define $k_{m,v}(n) \in \{0,1,\dots,K_m\}$ as the number of RBs allocated by BS-$m$ to vehicle-$v$ in the $n$-th time slot, and let $k_m(n) \in \{0,1,\dots,K_m\}$ represent the total number of RBs used by BS-$m$ in that time slot. BS-$m$ can allocate RBs to vehicle-$v$ only if a connection exists, hence
\begin{equation}
    k_{m,v}(n) = u_{m,v}(n)k_{m,v}(n).
    \label{eq:resource allocation cons}
\end{equation}
To avoid inter-user interference, a BS cannot assign an RB to multiple vehicles at a time, leading to the constraint
\begin{equation}
    k_{m}(n) = \sum_{v\in\mathcal{V}(n)} k_{m,v}(n) \leq K_m.
    \label{eq:RB cons}
\end{equation}
The transmission power of BS-$m$ in time slot-$n$ is based on its RB allocation and is given by
\begin{equation}
    P_m(n) = p_m k_{m}(n).
    \label{eq:BS instant power cons}
\end{equation}
The time-averaged transmission power of BS-$m$ is defined as
\begin{equation}
    \Bar{P}_m = \lim_{N\rightarrow\infty}\frac{1}{N}\sum_{n=0}^{N-1}P_m(n).
    \label{eq:BS timeavg power cons}
\end{equation}
Thus, the total time-averaged transmission power across all BSs in the vehicular network is
\begin{equation}
    \Bar{P} = \sum_{m=0}^M \Bar{P}_m =\lim_{N\rightarrow\infty}\frac{1}{N}\sum_{n=0}^{N-1} \sum_{m=0}^M \sum_{v\in\mathcal{V}(n)} p_m  k_{m,v}(n).
    \label{eq:area timeavg power cons}
\end{equation}

Define $G_{m,v}^{[x]}$ as the average channel gain between BS-$m$ and vehicle-$v$ in the $x$-th PHO frame. Assume that 
\begin{equation}
    G_{m,v}^{[x]} = \alpha_m B_{m,v}^{[x]} \left(d_{m,v}^{[x]}\right)^{-\beta_m} 10^{X_{m,v}^{[x]}/10},
    \label{eq:avg channel gain}
\end{equation}
where $\alpha_m$ is the power gain constant, $\beta_m$ is the path loss exponent, $d_{m,v}^{[x]}$ is the distance, $B_{m,v}^{[x]}$ represents the beamforming gain and $X_{m,v}^{[x]}$ captures the shadow fading effects. $X_{m,v}^{[x]}$ is assumed to be an independent and identically distributed (i.i.d.) zero-mean Gaussian random variable with variance $\sigma_X^2$. As beamforming is not the focus of this paper, we assume $B_{m,v}^{[x]} = \gamma_{\rm BF} N_{{\rm TX}}$, where $\gamma_{\rm BF} \in (0,1)$ represents the beamforming accuracy factor.
In the $i$-th time slot of the $x$-th PHO frame, the instantaneous channel gain $g_{m,v}^{[x]}(i)$ between BS-$m$ and vehicle-$v$ is modeled by the Rician fading channel as
\begin{equation}
    g_{m,v}^{[x]}(i) = G_{m,v}^{[x]}\left\|\sqrt{\frac{K_{\rm RIC}}{K_{\rm RIC}+1}} + \sqrt{\frac{1}{K_{\rm RIC}+1}} h_{m,v}^{[x]}(i)\right\|^2,
    \label{eq:instant channel condition}
\end{equation}
where $K_{\rm RIC}$ represents the Rician K-factor, $h_{m,v}^{[x]}(i) \sim \mathcal{CN}\left(0,1\right)$ is an i.i.d. standard complex Gaussian random variable.

Given that $k_{m,v}(n) > 0$, the signal-to-noise ratio (SNR) from BS-$m$ to vehicle $v$ is expressed as
\begin{equation}
    {\rm SNR}_{m,v}(n) = \frac{p_{m,v}(n) g_{m,v}(n)}{N_0 W_{m,v}(n)},
    \label{eq:gNB2V SNR}
\end{equation}
where $N_0$ denotes the noise power spectral density. The transmit power $p_{m,v}(n)$ and allocated bandwidth $W_{m,v}(n)$ from BS-$m$ to vehicle-$v$ are given by
\begin{align}
    p_{m,v}(n) & = k_{m,v}(n) p_m, \label{eq:transmit power}\\
    W_{m,v}(n) & = k_{m,v}(n) \Delta f_m. \label{eq:allocated bandwidth}
\end{align}
The transmission rate for downlink communication between BS-$m$ and vehicle-$v$ in time slot-$n$ is given by
\begin{align}
    r_{m,v}(n) & = W_{m,v}(n) \Delta t \log \left( 1 + {\rm SNR}_{m,v}(n) \right) \nonumber \\
    & = k_{m,v}(n) \Delta f_m \Delta t \log \left( 1 + \frac{ p_m g_{m,v}(n)}{N_0 \Delta f_m} \right).
    \label{eq:gNB2V rate origin}
\end{align}
Therefore, the maximum amount of information received by  vehicle-$v$ in time slot-$n$ is given by
\begin{equation}
    r_v(n) = \sum_{m=0}^M k_{m,v}(n) \Delta f_m \Delta t \log \left( 1 + \frac{ p_m g_{m,v}(n)}{N_0 \Delta f_m} \right).
    \label{eq:veh receive rate}
\end{equation}

\subsection{Traffic Model}
Define $q_v(n)$ as the data backlog queue length of vehicle-$v$ awaiting transmission by the BSs, which evolves as
\begin{equation}
    q_v(n+1) = \max \left\{ q_v(n) - r_v(n), 0 \right\} + a_v(n),
    \label{eq: queue update}
\end{equation}
where $a_v(n)$ is the amount of data arriving for vehicle-$v$ during time slot-$n$.
We assume that $a_v(n)$ is i.i.d. following a Poisson distribution with parameter $\lambda \Delta t$, where $\lambda$ is the average data arrival rate. According to \textit{Little's law} \cite{little1961proof}, the average queuing delay is proportional to the average queue length. This implies that the queuing delay $\tau_v(n)$ for vehicle-$v$ in time slot-$n$ can be approximated by $\tau_v(n) = \lambda^{-1}q_v(n)$. 
To meet the uRLLC QoS requirements \textcolor{black}{for V2X applications}, the queuing delay $\tau_v(n)$ should remain below the threshold $\tau^{{\rm th}}$, resulting in the QoS constraint
\begin{equation}
    q_v(n) \leq \lambda \tau^{{\rm th}}.
    \label{eq:queue length cons}
\end{equation}
Define the QoS violation probability as
\begin{equation}
    U = \lim_{N\rightarrow\infty}\frac{1}{N}\sum_{n=0}^{N-1}  \frac{\sum_{v\in\mathcal{V}(n)}U_v(n)}{V(n)},
    \label{eq:violation probability}
\end{equation}
where $U_v(n) = 1$ if $q_v(n) > \lambda \tau^{{\rm th}}$, and $U_v(n) = 0$ otherwise. The QoS violation probability $U$ serves as an indicator of system reliability in meeting the QoS constraint \eqref{eq:queue length cons}, following the approach in \cite{kien17}.

\section{Problem Formulation}{\label{sec:Problem Formulation}}

This paper aims to minimize the system energy consumption while adhering to QoS requirements and the constraints on HO and RA decisions. Specifically, the optimization problem is formulated as
\begin{equation}
    \label{prob1}
    \begin{aligned}
        \textbf{P1:} \quad  \min_{\bm{u}_{m,v},\bm{k}_{m,v}} & \Bar{P}  \\
        \textup{s.t.} \quad & \textup{Eqs. \eqref{eq:connection cons}-\eqref{eq:RB cons} and  \eqref{eq:queue length cons}} \\
        & u_{m,v}(n) \in \{0,1\},  \\
        & k_{m,v}(n) \in \{0,1,\dots,K_m\}.
    \end{aligned}
\end{equation}
The decision variables $\bm{u}_{m,v}$ and $\bm{k}_{m,v}$ are based on the HO and RA strategies, respectively. \Cref{{eq:connection cons},{eq:PHO cons}} represent the HO constraints, while \Cref{{eq:resource allocation cons},{eq:RB cons}} represent the RA constraints. The QoS requirement is captured by \Cref{eq:queue length cons}. 

Problem \textbf{P1} is an integer stochastic optimization problem, where the stochasticity stems from the factors such as vehicle arrivals and departures within the network area, changes in vehicle positions, random data traffic arrivals, and dynamic fluctuations in channel conditions. There are three key challenges for solving the problem:
\begin{enumerate}
    \item Different decision timescales: HO and RA decisions operate on different timescales. While $\bm{u}_{m,v}$ updates every $N_{{\rm HO}}$ time slots, $\bm{k}_{m,v}$ varies in each time slot, and $k_{m,v}(n)$ can be positive only if $u_{m,v}(n)=1$. 
    \item Environmental uncertainty: The randomness in traffic conditions, data arrivals, and channel states introduces uncertainty. Accurate future state information is difficult to obtain for making PHO and RA decisions.
    \item High computational complexity: Problem \textbf{P1} is an NP-hard integer programming problem with exponential complexity.
\end{enumerate}

Given that $\bm{u}_{m,v}$ and $\bm{k}_{m,v}$ operate on different timescales, problem \textbf{P1} is decomposed into sequential subproblems, each corresponding to a PHO frame. For each subproblem, the objective is to determine the optimal BS-vehicle connections and RB allocation within the PHO frame. The final state of each subproblem becomes the initial state for the next, forming a Markov decision process (MDP). The subproblem for the $x$-th PHO frame is formulated as
\begin{align}
\label{prob2}
\textbf{P2:} \qquad  \min_{u_{m,v}^{[x]},\bm{k}_{m,v}^{[x]}} & \frac{1}{N_{\textup{HO}}}\sum_{i=0}^{N_{\textup{HO}}-1} \sum_{m=0}^M \sum_{v\in\mathcal{V}^{[x]}} p_m  k_{m,v}^{[x]}(i) \\
\textup{s.t.} \quad 
& q_v^{[x]}(i) \leq \lambda \tau^{{\rm th}}, \forall i \in \mathcal{N}_{\textup{HO}},
\label{prob2: queue length cons} \\
& \sum_{m=0}^{M} u_{m,v}^{[x]}  \leq 1,
\label{prob2:connection cons} \\
& \sum_{v \in \mathcal{V}^{[x]}} k_{m,v} ^{[x]}(i) \leq K_m, \forall i \in \mathcal{N}_{\textup{HO}}, 
\label{prob2:subcarrier cons}\\
& k_{m,v}^{[x]}(i) = u_{m,v}^{[x]}k_{m,v}^{[x]}(i), \forall i \in \mathcal{N}_{\textup{HO}}, 
\label{prob2:resource allocation cons} \\
& u_{m,v}^{[x]} \in \{0,1\},
\label{prob2:u cons} \\
& k_{m,v}^{[x]}(i) \in \{0,1,\dots,K_m\}, \forall i \in \mathcal{N}_{\textup{HO}}.
\label{prob2:k cons}
\end{align}
Notably, in the PHO mechanism, the variable $u_{m,v}^{[x]}$ in problem \textbf{P2} for the $x$-th PHO frame must be determined in the $(x-1)$-th PHO frame. This ensures BSs complete signaling and resource reservations in advance for seamless HO. 
To support the PHO mechanism, an AI/ML-based method is introduced in Section \ref{subsection:AI} to predict future vehicle positions and channel states. Section \ref{subsection:PHO strategy} presents an energy-efficient PHO algorithm to efficiently approximate the optimal $u_{m,v}^{[x]}$.

Given the BS-vehicle connection states $u_{m,v}^{[x]}$ in the $x$-th PHO frame, problem \textbf{P2} can be decoupled into $M+1$ single-BS RA subproblems as
\begin{align}
\label{prob3}
\textbf{P3:} \qquad  \min_{\bm{k}_{m,v}^{[x]}} \quad & \frac{1}{N_{\textup{HO}}}\sum_{i=0}^{N_{\textup{HO}}-1} \sum_{v\in\mathcal{V}^{[x]}_m} p_m  k_{m,v}^{[x]}(i) \\
\textup{s.t.} \quad 
& q_v^{[x]}(i) \leq \lambda \tau^{{\rm th}}, \forall i \in \mathcal{N}_{\textup{HO}},
\label{prob3: queue length cons} \\
& \sum_{v \in \mathcal{V}^{[x]}_m} k_{m,v}^{[x]}(i) \leq K_m, \forall i \in \mathcal{N}_{\textup{HO}}, 
\label{prob3:subcarrier cons}\\
& k_{m,v}^{[x]}(i) = 0, \forall v \notin \mathcal{V}^{[x]}_m, \forall i \in \mathcal{N}_{\textup{HO}}, 
\label{prob3:resource allocation cons} \\
& k_{m,v}^{[x]}(i) \in \{0,1,\dots,K_m\}, \forall i \in \mathcal{N}_{\textup{HO}},
\label{prob3:k cons}
\end{align}
where $\mathcal{V}^{[x]}_m = \big{\{} v \in \mathcal{V}^{[x]} \big{|} u_{m,v}^{[x]}=1 \big{\}}$ represents the set of vehicles connected to BS-$m$ in the $x$-th PHO frame. Problem \textbf{P3} is still an integer stochastic optimization problem. Consequently, heuristic RA algorithms are proposed in Section \ref{subsection:RA strategy} to efficiently solve the problem.

\section{Algorithm Design}
In this section, we first present an AI/ML-based mobility prediction method to forecast the future positions and channel states for each vehicle. Subsequently, we propose energy-efficient PHO and heuristic RA strategies aimed at minimizing the system energy consumption while satisfying the QoS constraints. Finally, we introduce the AEPHORA framework, which integrates the aforementioned algorithms to facilitate energy-efficient PHO and RA decisions guided by the AI/ML-based prediction model.
 
\subsection{AI/ML-Based Prediction}{\label{subsection:AI}}

\begin{figure}
\centerline{\includegraphics[width=0.45\textwidth, trim= 150 120 300 100,clip]{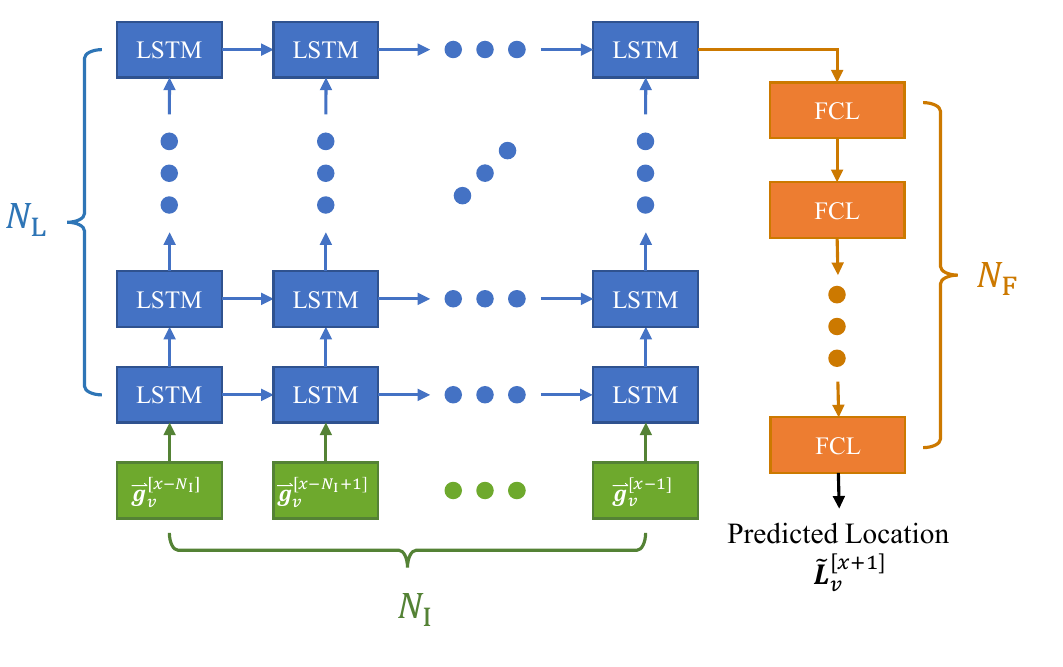}}
\caption{The AI/ML-based mobility prediction model.}
\label{fig:NN model}
\end{figure}

Figure \ref{fig:NN model} illustrates the architecture of the neural network (NN) model designed to predict the future location for each vehicle. The NN model comprises an input layer, $N_{\rm L}$ stacked Long Short-Term Memory (LSTM) layers and $N_{\rm F}$ stacked fully connected layers (FCLs) \cite{ozturk2019novel}. 
Each LSTM layer comprises 512 cells, while each FCL contains 64 neurons with ReLU activation, except for the final FCL, which has no activation function. Lower layers encode their inputs and pass the corresponding hidden states to subsequent layers. 
The model input consists of the historical channel gains between vehicle-$v$ and all BSs over the past $N_{\rm I}$ PHO frames, represented as $\bm{g}_v^{[x-1]}, \dots, \bm{g}_v^{[x-N_{\rm I}]} \in \mathbb{R}^{M+1}$, where $\bm{g}_v^{[x]} = \left[ g_{0,v}^{[x]}, \dots,g_{M,v}^{[x]} \right]^{\rm T}$.
The output is the predicted location $\Tilde{\bm{L}}_v^{[x+1]} \in \mathbb{R}^{2}$ for vehicle-$v$ in the next PHO frame. 
Denote the NN model as $F(\cdot;\bm{\theta})$, where $\bm{\theta}$ represents the model weights. Then we have
\begin{equation}
    \Tilde{\bm{L}}_v^{[x+1]} = F\left(\bm{g}_v^{[x-1]}, \dots, \bm{g}_v^{[x-N_{\rm I}]} ;\bm{\theta}\right).
    \label{eq:NN prediction}
\end{equation}

\subsection{Energy-Efficient PHO}{\label{subsection:PHO strategy}}

Given the predicted locations $\Tilde{\bm{L}}_v^{[x+1]}$, the expected average channel gain is estimated by 
\begin{equation}
    \Tilde{G}_{m,v}^{[x+1]} = \alpha_m \gamma_{\rm BF} N_{{\rm TX}} \left\| \Tilde{\bm{L}}_v^{[x+1]} - \bm{L}^{\textup{BS}}_m\right\|_2^{-\beta_m}.
    \label{eq:EPHO calculate distance}
\end{equation}
Subsequently, we can estimate the minimum average number of RBs required to meet the QoS constraint for vehicle-$v$ connected to BS-$m$ in the $(x+1)$-th frame, as follows:
\begin{equation}
    \Tilde{k}_{m,v}^{[x+1]} = \lambda \left[ \Delta f_m \log \left( 1 + \frac{ p_m \Tilde{G}_{m,v}^{[x+1]}}{N_0 \Delta f_m} \cdot 10^{-\frac{\sigma_X}{10}} \right) \right]^{-1}.
    \label{eq:EPHO estimate KB}
\end{equation}

By reserving $\Tilde{k}_{m,v}^{[x+1]}$ RBs for each vehicle-$v$ connected to BS-$m$ in the $(x+1)$-th frame while ensuring that the total reserved RBs of each BS-$m$ do not exceed its capacity $K_m$, the BS-vehicle association decision in problem \textbf{P2} can be approximated as a generalized assignment problem (GAP) as
\begin{align}
\label{prob4}
\textbf{P4:} \qquad  \min_{u_{m,v}^{[x+1]}} & \sum_{m=0}^M \sum_{v\in\mathcal{V}^{[x+1]}} p_m  \Tilde{k}_{m,v}^{[x+1]} u_{m,v}^{[x+1]} \\
\textup{s.t.} \quad 
& \sum_{v \in \mathcal{V}^{[x+1]}} \Tilde{k}_{m,v}^{[x+1]} u_{m,v}^{[x+1]} \leq K_m, \forall m \in \mathcal{M},
\label{prob4:subcarrier cons}\\
& \sum_{m=0}^{M} u_{m,v}^{[x+1]} = 1, \forall v\in\mathcal{V}^{[x+1]}
\label{prob4:connection cons} \\
& u_{m,v}^{[x+1]} \in \{0,1\}.
\label{prob4:u cons} 
\end{align}

Since GAP is NP-hard, we employ the polynomial-time approximation algorithm for GAP as presented in \cite{shmoys1993approximation} to develop an energy-efficient PHO schedule. This approximation algorithm can generate a schedule if and only if the linear programming (LP) relaxation of problem \textbf{P4} is feasible. In cases of significant network overload leading to infeasible LP relaxation, problem \textbf{P4} will be relaxed by repeatedly multiplying the right-hand side of constraint \eqref{prob4:subcarrier cons} by a relaxation factor $\zeta > 1$ until feasibility is achieved. In this paper, $\zeta$ is set to $1.1$. It has been proved in \cite{shmoys1993approximation} that if problem \textbf{P4} is feasible and its optimal solution yields an average power of $P^*$, the approximation algorithm in \cite{shmoys1993approximation} can generate a solution that results in an average power not exceeding $P^*$. 
However, the approximate solution only satisfies $\sum_{v \in \mathcal{V}^{[x+1]}} \Tilde{k}_{m,v}^{[x+1]} u_{m,v}^{[x+1]} \leq 2 K_m$ instead of constraint \eqref{prob4:subcarrier cons}, which may compromise the feasibility of the approximate solution.

The above strategy is called energy-efficient PHO (EPHO) algorithm, which is presented in \textbf{Alg. \ref{alg:EPHO}}.

\begin{algorithm}
    \caption{Energy-Efficient PHO (EPHO)}
    \label{alg:EPHO}
    \renewcommand{\algorithmicrequire}{\textbf{Input:}}
    \renewcommand{\algorithmicensure}{\textbf{Output:}}
    \begin{algorithmic}[1] 
        \REQUIRE  
        $\Tilde{\bm{L}}_v^{[x+1]}$, $\bm{L}^{\textup{BS}}_m$, $\alpha_m$, $\beta_m$, $p_m$, $\Delta f_m$, $\gamma_{\rm BF}$, $N_{{\rm TX}}$, $\zeta$, $N_0$;
        
        \ENSURE 
        $\hat{u}_{m,v}^{[x+1]}$;

        \STATE Calculate $\Tilde{k}_{m,v}^{[x+1]}$ by Eqs. \eqref{eq:EPHO calculate distance} and \eqref{eq:EPHO estimate KB};
        
        \WHILE{the LP relaxation of problem \textbf{P4} is not feasible}
            \STATE Multiply the right-hand side of constraint \eqref{prob4:subcarrier cons} by $\zeta$;
        \ENDWHILE
        
        \STATE Apply the GAP approximation algorithm in \cite{shmoys1993approximation} to solve problem \textbf{P4} to generate a PHO schedule $\hat{u}_{m,v}^{[x+1]}$;
        
        \STATE \textbf{return} 
    \end{algorithmic}
    
\end{algorithm}

\begin{algorithm}
    \caption{Heuristic RA (HRA)}
    \label{alg:HRA}
    \renewcommand{\algorithmicrequire}{\textbf{Input:}}
    \renewcommand{\algorithmicensure}{\textbf{Output:}}
    \begin{algorithmic}[1] 
        \REQUIRE  
        $u_{m,v}^{[x]}$, $g_{m,v}^{[x]}$, $q_v^{[x]}(i)$, $p_m$, $\Delta f_m$, $N_0$, $\lambda$, $\tau^{{\rm th}}$;
        
        \ENSURE 
        $k_{m,v}^{[x]}(i)$;

        Initialization:
        \STATE $k_{m,v}^{[x]}(i) \leftarrow 0$, $\Bar{K}_m \leftarrow K_m$, $\mathcal{V}^{[x]}_m \leftarrow \big{\{} v \in \mathcal{V}^{[x]} \big{|} u_{m,v}^{[x]}=1 \big{\}}$;
        
        \STATE Sort the vehicles in $\mathcal{V}^{[x]}_m$ by their queue lengths $q_v^{[x]}(i)$ in descending order to obtain a prioritized sequence $\bm{S}^{[x]}_m$.
        
        \FORALL{$v \in \bm{S}^{[x]}_m$}
            \STATE Calculate $\hat{k}_{m,v}^{[x]}(i)$ by Eqs. \eqref{eq:HRA estimate RB} and \eqref{eq:HRA final RB};
            
            \STATE $k_{m,v}^{[x]}(i) \leftarrow \hat{k}_{m,v}^{[x]}(i)$;
            
            \STATE $\Bar{K}_m \leftarrow \Bar{K}_m - k_{m,v}^{[x]}(i)$;
        \ENDFOR
        
        \STATE \textbf{return} 
    \end{algorithmic}
    
\end{algorithm}

\subsection{Heuristic RA}{\label{subsection:RA strategy}}
A heuristic RA (HRA) strategy is detailed in \textbf{Alg. \ref{alg:HRA}}, which guides each BS in allocating an appropriate number of RBs to connected vehicles in each time slot. The HRA strategy begins by sorting the connected vehicles in descending order based on their data backlog queue lengths. Subsequently, for each connected vehicle, the strategy estimates the average number of RBs needed to eliminate the accumulated backlogged data by
\begin{equation}
    \label{eq:HRA estimate RB}
    \Bar{k}_{m,v}^{[x]}(i) = q_v^{[x]}(i) \left[\Delta f_m \Delta t \log \left( 1 + \frac{ p_m g_{m,v}^{[x]}}{N_0 \Delta f_m} \right)\right]^{-1}.
\end{equation}
Since the allocated RB number for each vehicle must be an integer and their sum cannot exceed the RB capacity $K_m$, the final allocated RB number is given by
\begin{equation}
    \label{eq:HRA final RB}
    \hat{k}_{m,v}^{[x]}(i) = 
    \begin{cases}
        \min \left\{ \left\lceil\Bar{k}_{m,v}^{[x]}(i)\right\rceil, \Bar{K}_m \right\}, & \text{if } q_v^{[x]}(i) > \frac{\lambda \tau^{{\rm th}}}{2}, \\
        \min \left\{ \left\lfloor\Bar{k}_{m,v}^{[x]}(i)\right\rfloor, \Bar{K}_m \right\}, & \text{otherwise},
    \end{cases}
\end{equation}
where $\Bar{K}_m$ denotes the number of the remaining RBs at BS-$m$.

\subsection{AEPHORA}{\label{subsection:AEPHORA}}

\begin{algorithm}
    \caption{AI/ML-Based Energy-Efficient Proactive Handover and Resource Allocation (AEPHORA)}
    \label{alg:AEPHORA}
    \renewcommand{\algorithmicrequire}{\textbf{Input:}}
    \renewcommand{\algorithmicensure}{\textbf{Output:}}
    \begin{algorithmic}[1] 
        \REQUIRE  
        $\bm{\theta}$, $\mathcal{M}$, $N_0$, $N_{\textup{HO}}$, $\gamma_{\rm BF}$, $N_{{\rm TX}}$, $\zeta$, $\lambda$, $\Delta t$, $\tau^{{\rm th}}$, $K_m$, $p_m$, $\Delta f_m$, $\bm{L}^{\textup{BS}}_m$, $\alpha_m$, $\beta_m$, $\mathcal{V}^{[x]}$;
        
        \ENSURE 
        $u_{m,v}^{[x]}$,
        $k_{m,v}^{[x]}(i)$;

        Initialization:
        \STATE $\mathcal{V}^{[-1]} \leftarrow \emptyset$, $u_{m,v}^{[x]} \leftarrow 0$, $k_{m,v}^{[x]}(i) \leftarrow 0$;
        
        \FORALL{$x = 0,1,\cdots$}
            \STATE Newly arrived vehicles are initially connected to the macro BS: $u_{0,v}^{[x]} \leftarrow 1, \forall v \in \mathcal{V}^{[x]} \backslash \mathcal{V}^{[x-1]}$;
            
            \STATE Newly left vehicles handover to another macro BS outside the region $\mathcal{R}$: $u_{m,v}^{[x]} \leftarrow 0, \forall m \in \mathcal{M}, \forall v \in \mathcal{V}^{[x-1]} \backslash \mathcal{V}^{[x]}$;
            
            
            \STATE Each vehicle measures and reports the instantaneous channel gains $g_{m,v}^{[x]}, \forall m \in \mathcal{M}, \forall v \in \mathcal{V}^{[x]}$;
            
            \STATE Based on historical channel state information, utilize the AI/ML-based prediction method to forecast the positions $\Tilde{\bm{L}}^{[x+1]}_v$ of each vehicle in the next PHO frame, $\forall v \in \mathcal{V}^{[x]}$;
            
            \STATE \textbf{Algorithm \ref{alg:EPHO}} is applied to make the PHO decision $\hat{u}_{m,v}^{[x+1]}$ and initiate the PHO process for the next PHO frame;
            
            \FORALL {$i=0,1,\cdots,N_{\textup{HO}}-1$}
                \FORALL {$m \in \mathcal{M}$}
                    \STATE \textbf{Algorithm \ref{alg:HRA}} is applied for the RA decision $k_{m,v}^{[x]}(i)$;
                \ENDFOR
            
                
                \STATE Each BS conducts downlink communication and updates the data backlog queue $q_v^{[x]}(i+1)$ for all the connected vehicles based on Eqs. \eqref{eq:veh receive rate} and \eqref{eq: queue update};
            \ENDFOR
            \STATE Complete the PHO process for the next PHO frame and update the connection states accordingly: $u_{m,v}^{[x+1]} \leftarrow \hat{u}_{m,v}^{[x+1]}$, $\forall m \in \mathcal{M}, \forall v \in \mathcal{V}^{[x]}$;
        \ENDFOR
        \STATE \textbf{return} 
    \end{algorithmic}
\end{algorithm}

\begin{figure}
\centerline{\includegraphics[width=0.45\textwidth, trim= 0 10 430 10,clip]{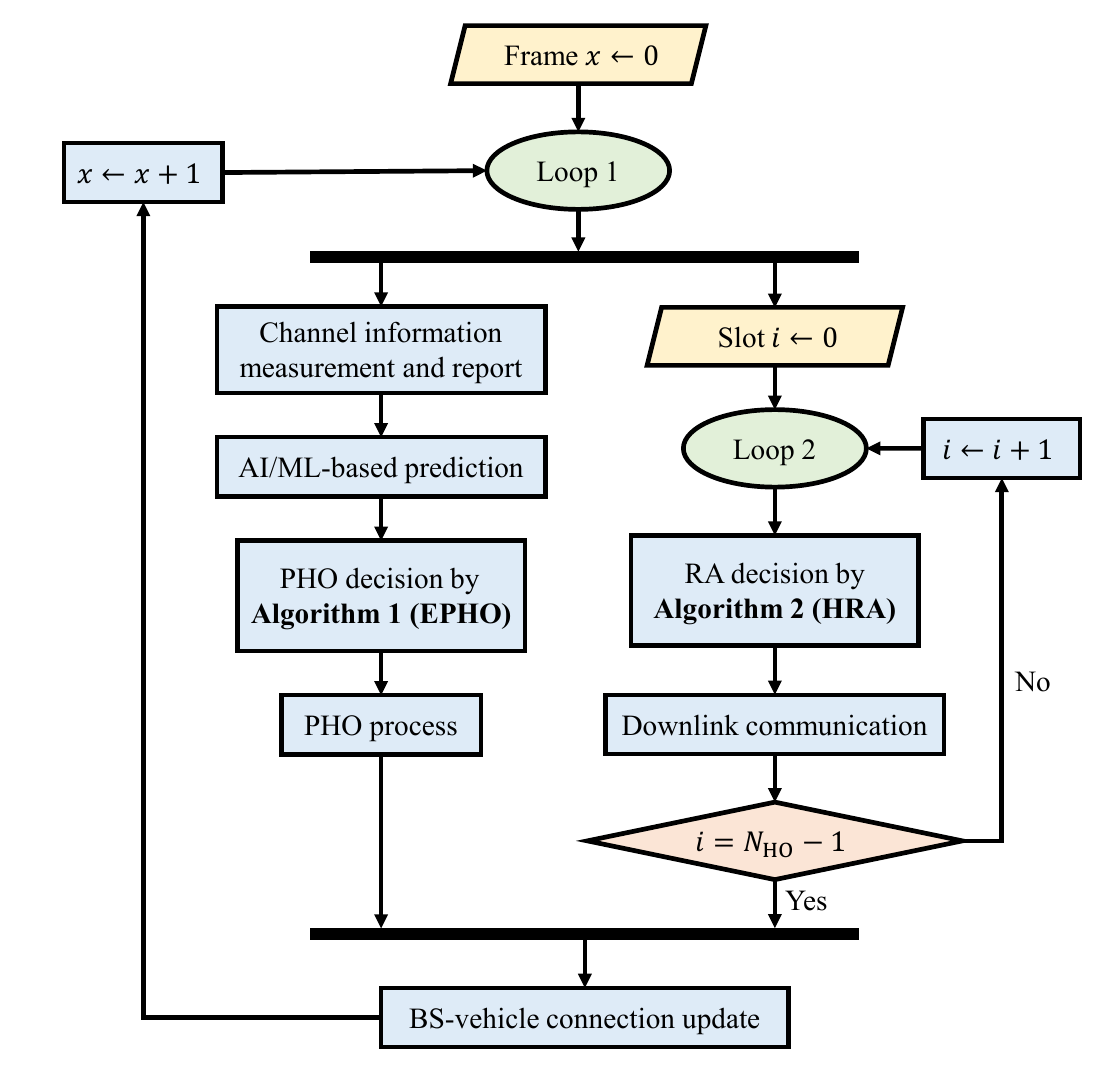}}
\caption{Flowchart of the AEPHORA algorithm.}
\label{fig:flow chart}
\end{figure}

\textbf{Algorithm \ref{alg:AEPHORA}} demonstrates the AI/ML-based energy-efficient PHO and RA optimization framework (\textbf{AEPHORA}). Within each PHO frame, newly arrived vehicles are initially connected to the macro BS, while those departing the region are disconnected. Each vehicle measures and reports its instantaneous channel gains, providing real-time data for decision-making. Utilizing historical channel state information, the AEPHORA framework applies the AI/ML-based prediction strategy proposed in Section \ref{subsection:AI} to forecast future vehicle positions, enabling PHO management. PHO decisions are derived from the outcomes of \textbf{Alg. \ref{alg:EPHO}}, while RA decisions are determined using \textbf{Alg. \ref{alg:HRA}}. During each time slot in the current PHO frame, each BS conducts downlink communication and updates the data backlog queues for all connected vehicles. At the end of the current PHO frame, the PHO process for the subsequent frame is completed, and the BS-vehicle connection states are updated accordingly. We emphasize that the PHO process and the RA process occur in parallel, as illustrated in the flowchart of \textbf{Algorithm \ref{alg:AEPHORA}} in Fig. \ref{fig:flow chart}. 

\section{Simulation Results}{\label{sec:Simulation Results}}

In this section, simulation results demonstrate the effectiveness of the proposed AEPHORA framework. The simulation scenario aligns with Figure \ref{fig:system model}, featuring one macro BS and four micro BSs positioned within an 800 m by 800 m area. The macro BS is located at the area center $(0 \text{ m}, 0 \text{ m})$, while the coordinates of the four micro BSs are $(\pm 250 \text{ m}, \pm 250 \text{ m})$. The centers of the four intersections are situated at $(\pm 200 \text{ m}, \pm 200 \text{ m})$. Each road consists of two directions, with three lanes allocated for each direction. The vehicular trajectories are generated using the road traffic simulator SUMO \cite{behrisch2011sumo}, providing realistic traffic dynamics for the simulations. 
The average number of vehicles in the considered region is about 220, with an average of 1.6 vehicles entering and 1.6 vehicles leaving the region per second.
The channel gains for each vehicle and BS are randomly generated according to Eqs. \eqref{eq:avg channel gain} and \eqref{eq:instant channel condition}. The parameter settings for simulation are summarized in Table \ref{tab:simulation parameter}.

\begin{table}[ht]
    \centering
    \caption{Simulation Parameter Settings}
    \begin{tabular}{c|cc||c|cc}
        \toprule
        \multirow{2}{*}{\textbf{Parameter}} & \multicolumn{2}{c||}{\textbf{Value}} & \multirow{2}{*}{\textbf{Parameter}} & \multirow{2}{*}{\textbf{Value}} \\ \cline{2-3} & \textbf{Macro BS} & \textbf{Micro BS} & & \\ \midrule
        $\alpha_m$ &  -55.4 dB  &  -72.0 dB & $\sigma_X$ &  {4} \\ 
        $\beta_m$ &  2.2  &  3 & $\Delta t$ &  {1 ms} \\
        $f_m$ &  5.9 GHz  & 24 GHz & $\tau^{{\rm th}}$ &  {20 ms}  \\
        $\Delta f_m$ &  0.18 MHz  & 1.8 MHz & $K_{\rm RIC}$ &  {10}  \\
        $K_m$ &  100  & 100  & $N_{\rm HO}$ &  {100}  \\
        $p_m$ &  30 dBm  & 20 dBm  & $N_{{\rm TX}}$ &  {32}  \\
        $N_0$ &  \multicolumn{2}{c||}{-173 dBm/Hz} & $\gamma_{\rm BF}$ &  {0.5}  \\
        \bottomrule
        \multicolumn{1}{c|}{\multirow{3}{*}{\textbf{Others}}} & \multicolumn{4}{c}{\multirow{3}{*}{\makecell{Simulation Duration = 30 s \\ 
        Maximum driving velocity = 18.34 m/s \\ Average data arrival rates in [1.0,\,1.5,\,2.0,\,\dots,\,13.0] Mbps}}}\\
        \\ \\ \bottomrule
        
    \end{tabular}
    \label{tab:simulation parameter}
\end{table}

\begin{table}[ht]
    \centering
    \caption{Model Training Parameters and Validation Results}
    \begin{tabular}{c|c|c|c|c}
        \toprule
        \multirow{2}{*}{\textbf{$N_{\rm L}$}} & \multirow{2}{*}{\textbf{$N_{\rm F}$}} &  \multicolumn{2}{c|}{\textbf{Validation MAE}} & \multirow{2}{*}{\textbf{Training Parameters}} \\ \cline{3-4} 
        &&$\sigma_X=0$&$\sigma_X=4$&\\   \midrule 
         \multirow{3}{*}{1}  &  1 & 2.967 m &  18.925 m & \multirow{12}{*}{
           \makecell{ {Training Dataset Size} = 1608415 \\{Validation Dataset Size} = 643366 \\{Epoch Number} = 100 \\ {Learning Rate} = $10^{-3}$ \\ {Weight Decay} = $10^{-5}$ \\ {Batch Size} = 64  \\ $N_{\rm I} = 10$}} \\ \cline{2-4}
          &  2 & 2.988 m & 18.705 m & \\ \cline{2-4}
          &  3 & \textbf{2.965 m} & 18.703 m & \\ \cline{1-4}
          \multirow{3}{*}{2}  &  1 & 2.971 m &  18.791 m & \\ \cline{2-4}
          &  2 & 2.985 m & \textbf{18.668 m} & \\ \cline{2-4}
          &  3 & 2.980 m & 18.714 m & \\ \cline{1-4}
          \multirow{3}{*}{3}  &  1 & 2.975 m &  18.829 m & \\ \cline{2-4}
          &  2 & 3.033 m & 18.851 m & \\ \cline{2-4}
          &  3 & 2.999 m & 18.913 m & \\ \cline{1-4}
        \multirow{3}{*}{4}  &  1 & 3.054 m &  18.849 m & \\ \cline{2-4}
          &  2 & 3.063 m & 18.852 m & \\ \cline{2-4}
          &  3 & 3.183 m & 19.073 m & \\ 
        \bottomrule
    \end{tabular}
    \label{tab:NN training and validation}
\end{table}

We train the NN model using the Adam optimizer to minimize the mean squared error (MSE) between the predicted location $\Tilde{\bm{L}}_v^{[x+1]}$ and the true location ${\bm{L}}_v^{[x+1]}$. The performance of the trained models is validated using the mean absolute error (MAE) as a key performance indicator. The parameter settings for model training, along with the results of the performance validation, are summarized in Table \ref{tab:NN training and validation}. Increasing $N_{\rm L}$ and $N_{\rm F}$ does not guarantee improved performance. The optimal model parameters for the realistic environment with $\sigma_X = 4$ are $N_{\rm L} = 2 $ and $N_{\rm F} = 2$, which will be utilized in the simulations.

We compare the performance of the proposed \text{AEPHORA} scheme with three other baseline schemes listed as follows:
\begin{enumerate}
    \item \textbf{HEE}: Each vehicle connects to the BS with the highest energy efficiency indicator, defined as the predicted transmission rate per unit of transmission power: $\frac{\Delta f_m}{p_m}\log\left( 1 + p_m \frac{\Tilde{G}_{m,v}^{[x+1]}}{N_o \Delta f_m} \right)$. The RA strategy is \textbf{Alg. \ref{alg:HRA}}.
    \item \textbf{HRBE}: Each vehicle connects to the BS with the highest RB efficiency indicator, defined as the predicted transmission rate per RB: $\Delta f_m\log\left( 1 + p_m \frac{\Tilde{G}_{m,v}^{[x+1]}}{N_o \Delta f_m} \right)$. The RA strategy is \textbf{Alg. \ref{alg:HRA}}.
    \item \textbf{InfRB\_LB}: The PHO strategy aligns with \text{HEE}, but without considering integer and capacity constraints for RB allocation. Each vehicle is allocated $\Bar{k}_{m,v}^{[x]}(i)$ RBs. While this RA strategy cannot be directly applied due to practical limitations, it establishes a lower bound on energy consumption necessary to satisfy the QoS constraint \eqref{eq:queue length cons}.
\end{enumerate}
Additionally, to assess the impact of the AI/ML-based prediction model on the system performance, each scheme is further divided into two versions: one employs vehicle positions predicted by the NN model, while the other utilizes actual vehicle positions. 

\begin{figure}[t]
\centering
    \subfloat[Average system power $\Bar{P}$ versus average data arrival rate $\lambda$.]    {\includegraphics[width=0.9\columnwidth,trim= 10 5 10 20,clip]{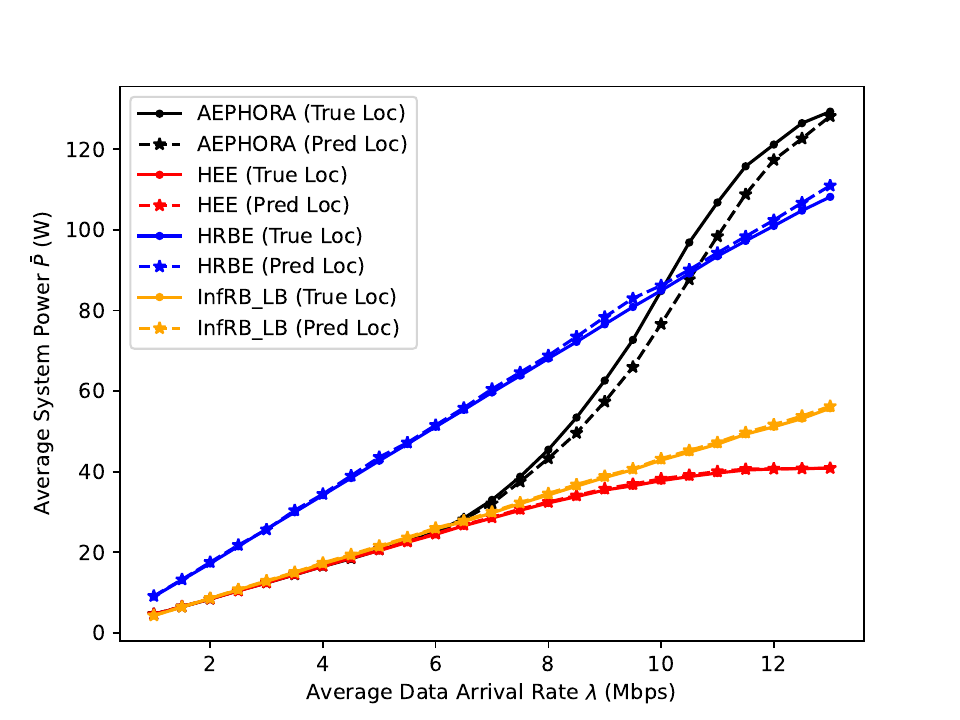}\label{fig:Average system power}}    \\
    \subfloat[QoS violation probability $U$ versus average data arrival rate $\lambda$.]    {\includegraphics[width=0.9\columnwidth,trim= 10 5 10 20,clip]{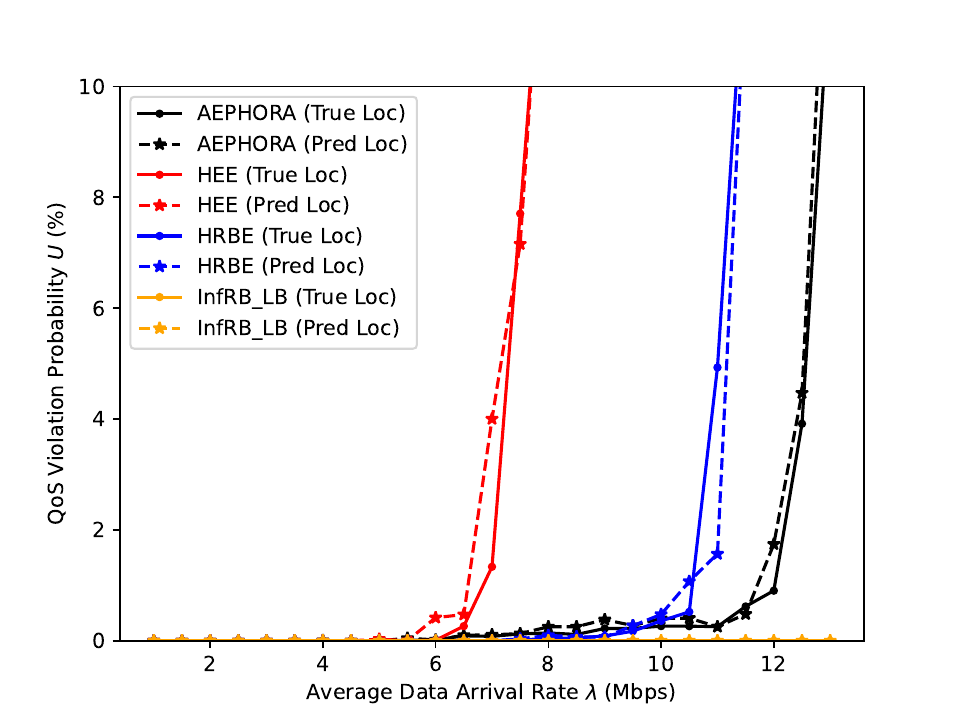}\label{fig:Violation probability}}
\caption{Comparison between the proposed AEPHORA and other baseline schemes.}
\label{fig:scheme comparison}
\end{figure}

The average system power $\Bar{P}$ versus average data arrival rate $\lambda$ of these schemes is illustrated in Fig. \ref{fig:Average system power}, while the QoS violation probability $U$ is presented in Fig. \ref{fig:Violation probability}. Solid curves with point markers represent the schemes utilizing actual vehicle positions, while dashed curves with star markers correspond to the schemes employing NN-predicted locations. The solid and dashed lines for the same scheme are closely aligned, indicating that the algorithm performance loss due to NN prediction errors is acceptable.

In Figure \ref{fig:scheme comparison}, the simulation results can be divided into four stages based on the value of $\lambda$:
\begin{enumerate}
    \item $\lambda \leq 6 \text{ Mbps}$: For all these four schemes, $U$ is nearly zero, with $\Bar{P}$ of \text{AEPHORA} and \text{HEE} closely aligning with the lower bound baseline {InfRB\_LB}. \text{HRBE} exhibits a significantly higher $\Bar{P}$ due to its prioritization of vehicular connections to the BS requiring the fewest RBs, thus sacrificing energy efficiency for increased system capacity.
    
    \item $6 \text{ Mbps} < \lambda \leq 10 \text{ Mbps}$: $\Bar{P}$ of \text{AEPHORA} rises more rapidly, approaching that of \text{HRBE} around $\lambda = 10 \text{ Mbps}$. This indicates that \text{AEPHORA} effectively balances the load by the adaptive handover of vehicles from overloaded micro BSs to the macro BS. While this results in an increase in $\Bar{P}$ , it ensures that $U$ remains sufficiently low. In contrast, $U$ for \text{HEE} rises sharply, and its $\Bar{P}$ drops below that of {InfRB\_LB} because it fails to offload vehicles from overloaded micro BSs to the macro BS, resulting in inadequate transmission rates.
    
    \item $10 \text{ Mbps} < \lambda \leq 11.5 \text{ Mbps}$: \text{AEPHORA} surpasses \text{HRBE} in $\Bar{P}$. However, $U$ for \text{AEPHORA} remains near zero, while that for \text{HRBE} rises sharply. This suggests that \text{AEPHORA} maintains better uRLLC QoS through adaptive load balancing under high data rates compared to \text{HRBE}.
    
    \item $\lambda > 11.5 \text{ Mbps}$: $U$ for \text{AEPHORA} increases sharply, indicating that the traffic load exceeds the system capacity.
\end{enumerate}

Specifically, \text{AEPHORA} supports a maximum data arrival rate with the QoS violation probability close to zero, nearly double that of \text{HEE}. Additionally, \text{AEPHORA} consumes only about half the system power of \text{HRBE} at low data arrival rates, while supporting higher maximum data arrival rates than \text{HRBE}.

\section{Conclusion}
The proposed \text{AEPHORA} framework effectively balances system energy consumption with uRLLC QoS requirements, providing a significant advantage over the practical baselines \text{HEE} and \text{HRBE}. Under light traffic loads, \text{AEPHORA} approximates the lower bound of system power.  With heavy traffic loads, \text{AEPHORA} adaptively offloads vehicles from overloaded micro BSs to the macro BS, ensuring QoS requirements at the cost of a manageable increase in power consumption. Consequently, \text{AEPHORA} achieves lower energy consumption while supporting heavier traffic loads, demonstrating its efficiency in real-world applications. However, the performance of \text{AEPHORA} depends on the accuracy of AI/ML-based predictions, as prediction errors can result in sub-optimal handover decisions. Better AI/ML-based mobility prediction models are required to further optimize the PHO process of \text{AEPHORA}.


\bibliographystyle{ieeetr}
\bibliography{main.bib}

\end{document}